\input harvmac.tex 

\def\Asym#1#2{\vcenter{\vbox{\drawbox{#1}{#2}
              \kern-#2pt       
              \drawbox{#1}{#2}}}}

\Title{\vbox{\rightline{hep-th/9807090} \rightline{CERN-TH/98-229}
}}
{\vbox{\centerline{Bulk Gauge Fields in AdS Supergravity}
\centerline{and Supersingletons}}}


\centerline{Sergio Ferrara} \smallskip{\it
\centerline{CERN Geneva, Switzerland}}
\centerline{\tt
Sergio.Ferrara@cern.ch}

\vskip .3cm
\centerline{Alberto Zaffaroni} \smallskip{\it
\centerline{CERN Geneva, Switzerland}}
\centerline{\tt
Alberto.Zaffaroni@cern.ch}

\vskip .1in


\noindent
We describe conformal operators living at the boundary of $AdS_{d+1}$ in a general setting. Primary conformal operators at the threshold of the unitarity bounds of UIR's of O(d,2) correspond to singletons and massless fields in $AdS_{d+1}$, respectively.
For maximal supersymmetric theories in $AdS_{d+1}$ we describe ``chiral'' primary short supermultiplets and non-chiral primary long supermultiplets. Examples are exhibited which correspond to KK and string states. We give the general contribution of a primary conformal operator to the OPE and Green's functions of primary fields, which may be relevant to compute string corrections to the four-point supergraviton amplitude in Anti-de-Sitter space.
\vskip 2.5truecm
{\it The material in this paper was presented by one of the authors (S.F) at
String 98, Santa Barbara.}

\noindent
CERN-TH/98-229
\Date{July 98}

\lref\malda{ J. M. Maldacena, {\it The Large N Limit of Superconformal Field Theories and Supergravity},  hep-th/9705104.}
\lref\maldatwo{ N. Itzhaki, J. M. Maldacena, J. Sonnenschein and S. Yankielowicz, {\it Supergravity and The Large N Limit of Theories With Sixteen Supercharges}, hep-th/9802042.}
\lref\witten{E. Witten, {\it Anti-de Sitter Space And Holography}, hep-th/9802150.}
\lref\pol{S. S. Gubser, I. R. Klebanov and A. M. Polyakov, {\it Gauge Theory Correlators from Non-Critical String Theory}, hep-th/9802109.}
\lref\fer{S. Ferrara and  C. Fr\o nsdal, {\it  Conformal Maxwell theory as a singleton field theory on AdS$_5$, IIB three branes
     and duality}, hep-th/971223;{\it Gauge Fields as Composite Boundary Excitations}, hep-th/9802126.}
\lref\fz{S. Ferrara and B. Zumino, Nucl. Phys. B87 (1975) 207.}
\lref\fertwo{S. Ferrara and  C. Fr\o nsdal, {\it Gauge Fields as Composite Boundary Excitations}, hep-th/9802126.}
\lref\van{H. J. Kim, L. J. Romans and P. van Nieuwenhuizen, Phys. Rev. D23 (1981) 1278.}
\lref\roo{E. Bergshoeff, M. De Roo and B de Wit, Nucl. Phys. B182 (1981) 173.}
\lref\stelle{P. Howe, K. S. Stelle and P. K. Townsend, Nucl. Phys. B192 (1981) 332.}
\lref\konishi{K. Konishi, Phys. Lett. B135 (1984) 439.}
\lref\kleb{C. W. Gibbons and P. K. Townsend, Phys. Rev. Lett. 71 (1993) 3754; M. P. Blencowe and M. J. Duff, Phys. Lett. B203 (1988) 229; Nucl. Phys B310 (1988), 389; M. J. Duff, Class. Quantum Grav. 5 (1988) 189; E. Bergshoeff, M. J. Duff, C. N. Pope and E. Sezgin, Phys. Lett. B199 (1988) 69; H. Nicolai, E. Sezgin and Y. Tanii, Nucl. Phys B305 (1988) 483.}
\lref\sken{K. Sfetsos and K. Skenderis, {\it Microscopic derivation of the Bekenstein-Hawking entropy formula for
     non-extremal black holes}, hep-th/9711138.}
\lref\skentwo{H. J. Boonstra, B. Peeters and K. Skenderis, {\it Branes and anti-de Sitter spacetimes}, hep-th/9801076.}
\lref\kall{P. Claus, R. Kallosh and A. Van Proeyen, {\it M 5-brane and superconformal (0,2) tensor multiplet in 6 dimensions}, hep-th/9711161.}
\lref\kalltwo{ R. Kallosh, J. Kumar and  A. Rajaraman, {\it Special Conformal Symmetry of Worldvolume Actions}, hep-th/9712073.}
\lref\kallthree{P. Claus, R. Kallosh, J. Kumar, P. Townsend and A. Van Proeyen, {\it Conformal theory of M2, D3, M5 and D1+D5 branes}, hep-th/9801206.}
\lref\ooguri{G. T. Horowitz and H. Ooguri, {\it Spectrum of Large N Gauge Theory from Supergravity}, hep-th/9802116.}
\lref\silv{S. Kachru and  E. Silverstein, {\it 4d Conformal Field Theories and Strings on Orbifolds}, hep-th/9802183.}
\lref\guntwo{M. Gunaydin and N. Marcus, Class. Quantum Grav. 2 (1985) L11.}
\lref\fztwo{S. Ferrara and B. Zumino, Nucl. Phys. B134 (1978) 301.}
\lref\gun{M. Gunaydin, L. J. Romans and N. P. Warner, Phys. Lett. 154B (1985) 268; M. Pernici, K. Pilch and P. van Nieuwenhuizen, Nucl. Phys. B259 (1985) 460.}
\lref\fr{M. Flato and C. Fr\o nsdal, J. Math. Phys. 22 (1981) 1100; Phys. Lett. B172 (1986) 412.}
\lref\frtwo{M. Flato and C. Fr\o nsdal, Lett. Math. Phys 2 (1978) 421; Phys. Lett. 97B (1980) 236.}
\lref\frthree{E. Angelopoulos, M. Flato, C. Fr\o nsdal and D. Sternheimer, Phys. Rev. D23 (1981) 1278.}
\lref\anselmi{D. Anselmi, M. Grisaru, A. Johansen, Nucl.Phys. B491 (1997) 221;
 D. Anselmi, D. Z. Freedman, M. T. Grisaru, A. A. Johansen, Phys. Lett. B394 (1997) 329}
\lref\jack{I. Jack, Nucl. Phys. B253 (1985) 323.}
\lref\jones{For a recent discussion, see I. Jack, T. D. Jones and C. G. North,
Phys. Lett. B386 (1996) 13.} 
\lref\freed{D. Z. Freedman, S. D. Mathur, A. Matusis, L. Rastelli, {\it Correlation functions in the CFT(d)/AdS(d+1) correpondence}, hep-th/9804058.}
\lref\tse{H. Liu and A. A. Tseytlin, {\it D=4 Super Yang Mills, D=5 gauged supergravity and D=4 conformal supergravity}, hep-th/9804083.}
\lref\seib{S. Lee, S. Minwalla, M. Rangamani and N. Seiberg, {\it Three-Point Functions of Chiral Operators in D=4, N=4 SYM at Large N}, hep-th/9806074.}
\lref\FFZ{S. Ferrara, C. Fr\o nsdal and A. Zaffaroni, {\it On N=8 Supergravity on $AdS_5$ and N=4 Superconformal Yang-Mills theory }, hep-th/9802203.}
\lref\grillo{S. Ferrara, A. F. Grillo and R. Gatto, Ann. Phys. 76 (1973) 161.}
\lref\grillotwo{S. Ferrara, A. F. Grillo and R. Gatto, Lett. Nuovo Cimento 2 (1971) 1363. S. Ferrara, A. F. Grillo, R. Gatto and G. Parisi, Nucl. Phys. B49 (1972) 77; Nuovo Cimento 19 (1974) 667.}
\lref\FA{L. Andrianopoli and S. Ferrara, {\it K-K excitations on AdS$_5\times S^5$ as N=4 ``primary'' superfields}, hep-th/9803171.}
\lref\binegar{B. Binegar, C. Fr\o nsdal and W. Heidenreich, J. Math. Phys. 24 (1983) 2828.}
\lref\fggp{S. Ferrara, R. Gatto and  A. F. Grillo, Springer Tracts in Modern Physics, vol. 67 (Berlin-Heidelberg), Springer, New York,  (1973).}
\lref\fggptwo{S. Ferrara, R. Gatto, A. F. Grillo and G. Parisi, in {\it Scale and Conformal symmetry in hadron physics}, edited by R. Gatto, Wiley, New York, (1983).}
\lref\luscher{G. Mack and M. Luscher, Comm. Math. Phys. 41 (1975) 203.}
\lref\macktwo{G. Mack, J. Phys. 34, Colloque C-1 (Suppl. au no. 10) (1973) 79.}
\lref\mackthree{G. Mack and I. Todorov, Phys. Rev. D8 (1973) 1764.}
\lref\mack{G. Mack and A. Salam, Ann. Phys. 53 (1969) 174.}
\lref\fe{S. Ferrara, Nucl. Phys. B77 (1974) 73.}
\lref\fgg{S. Ferrara, A. F. Grillo and R. Gatto, Ann. Phys. 76 (1973) 161; Phys. Rev. D9, (1974) 3564.} 
\lref\gunaydin{M. Gunaydin, D. Minic and  M. Zagermann, {\it 4D Doubleton Conformal Theories, CPT and IIB String on AdS$_5\times S^5$}, hep-th/9806042.}
\lref\dirac{P. A. M. Dirac, Ann. Math. 37 (1936), 429.}
\lref\nic{D. Z. Freedman and H. Nicolai, Nucl. Phys. B237 (1984) 342.}
\lref\howe{P. S. Howe and  P. C. West, {\it Non-perturbative Green's functions in theories with extended superconformal
     symmetry}, hep-th/9509140; Phys. Lett. B389 (1996) 273; {\it Is N=4 Yang-Mills Theory Soluble?}, hep-th/9611074; Phys.Lett. B400 (1997) 307.}
\lref\FZL{S. Ferrara, M. LLedo and A. Zaffaroni, {\it Born-Infeld Corrections to D3 brane Action in $AdS_5\times S_5$ and N=4, d=4 Primary
     Superfields}, hep-th/9805082.}
\lref\KL{I. R. Klebanov, Nucl.Phys. B496 (1997) 231.}
\lref\das{S. R. Das and S. P. Trivedi, {\it Three Brane Action and The Correspondence Between N=4 Yang Mills Theory and Anti De
     Sitter Space}, hep-th/9804149.}
\lref\banks{T. Banks and M. B. Green, JHEP05(1988)002.}
Some aspects of the AdS$_{d+1}$/CFT correspondence, inspired by the original conjecture by Maldacena\malda, and subsequently sharpened by Gubser, Klebanov, Polyakov and Witten \refs{\pol,\witten}, are considered in this paper\foot{For earlier work on the connection between branes geometry, Anti-de-Sitter space and singletons, see \kleb}. The maximal supersymmetric case, corresponding to the horizon geometry of D3 branes in type IIB strings, is considered as basic example.

More specifically, we describe the relation between interacting non-abelian singletons on $\partial$AdS$_5$, i.e. N=4 SU(n) super Yang-Mills theory and the bulk supergravity theory on AdS$_5\times S^5$.

In section 1, we review some properties of primary conformal fields on $\partial$AdS$_5$ and, in section 2, their superextension.
In sections 3 and 4, we describe several types of N=4 multiplets, realized in N=4 super Yang-Mills theory, which in the AdS/CFT correspondence, are related to singletons, massless and massive KK chiral primary supermultiplets as well as massive higher spin multiplets. In section 5, we briefly describe manifestly
covariant OPE on $\partial$AdS$_5$ which can be used to compute boundary correlation functions of primary operators corresponding to states in the spectrum of the type IIB string compactified on AdS$_5\times S^5$. In particular, the contribution to the four-point function of chiral primaries, due to the non-chiral primaries corresponding to string states, is exhibited. The latter may be used to connect string corrections to the four-point graviton amplitude in AdS$_5\times S^5$ geometry to four-point stress-energy tensor correlation functions on the boundary.

\newsec{Conformal fields on $\partial AdS_5$}
The 15 generators $J_{AB}=-J_{BA}$ of the conformal algebra O(4,2) can be defined in terms of the Poincar\'e generators $P_\mu ,M_{\mu\nu}$, the dilatation  $D$ and the special conformal transformation $K_\mu$, by the relations,
\eqn\algebra{J_{\mu\nu}=M_{\mu\nu},\qquad J_{5\mu}={1\over 2}(P_\mu -K_\mu ),\qquad J_{6\mu}={1\over 2}(P_\mu + K_\mu ),\qquad
J_{65}=D.}
The irreducible representations of the conformal algebra are specified by the values of the three Casimir operators \mack ,
\eqn\casimir{\eqalign{C_I&=J^{AB}J_{AB}=\cr C_{II}&=\epsilon_{ABCDEF}J^{AB}J^{CD}J^{EF}\cr C_{III}&=J_A^BJ_B^CJ_C^DJ_D^A}}
Every irreducible (infinite dimensional) representation can be specified by an
irreducible representation of the Lorentz group with definite conformal dimension and annihilated by $K_\mu$. In terms of the stability algebra at $x=0$ ($K_\mu,D, M_{\mu\nu}$), we define a primary conformal field, in a generic representation of the Lorentz group, by
\eqn\conf{\eqalign{[O_{\{\alpha\}}(0),D]&=ilO_{\{\alpha\}}(0)\cr
 [O_{\{\alpha\}}(0),K_\mu ]&=0.}}
The descendents $\partial ....\partial O_{\{\alpha\}}(0)$ fill an infinite dimensional representation specified by three numbers $(l,j_L,j_R)$, which contain
the conformal dimension and the Lorentz quantum number of the primary operator. 

In terms of the three Casimirs of the stability algebra, 
\eqn\casimirtwo{D=l,\, {1\over 2}M_{\mu\nu}M^{\mu\nu}=j_L(j_L+1)+j_R(j_R+1), \,
{1\over 2}\epsilon_{\mu\nu\rho\sigma}M^{\mu\nu}M^{\rho\sigma}=j_L(j_L+1)-j_R(j_R+1)}
the O(4,2) Casimirs take the following values\gunaydin : 

\eqn\tensorone{\eqalign{C_I&=l(l-4)+2j_L(j_L+1)+2j_R(j_R+1)\cr C_{II}&=(l-2)(j_L(j_L+1)-j_R(j_R+1))\cr C_{III}&=(l-2)^4-4(l-2)^2[j_L(j_L+1)+j_R(j_R+1)+1] +16j_Lj_R(j_L+1)(j_R+1)}}

In the particular case of  a tensor representation of spin s $(l,j_L=j_R=s/2)$, we have:
\eqn\tensor{\eqalign{C_I&=l(l-4)+s(s+2)\cr C_{II}&=0\cr C_{III}&=[l(l-2)-s(s+2)][(l-2)(l-4)-s(s+2)]}}
Let us consider, for example, the case of a conformal scalar. It is convenient
to consider a six-dimensional space with signature $(+---,-+)$ where the O(4,2)
generators act on the coordinates $\eta_A, A=0,...,5$ as $L_{AB}=i(\eta_A\partial_B-\eta_B\partial_A)$ \refs{\dirac,\mack}. The conformal compactification of Minkowski space-time can be identified with the hypercone $\eta_A\eta^A=0$, with projectively identified coordinates ($\eta^A=\lambda\eta^A$). A conformal scalar can be represented as a homogeneous function on the hypercone $\eta_A\eta^A=0$ \refs{\dirac,\mack},
\eqn\scal{(\eta\partial )\Phi=\lambda\Phi,\qquad\qquad\qquad (\lambda =-l)}
The quadratic Casimir
\eqn\quad{{1\over 2}L_{AB}L^{AB}=-\eta^2\partial^2 +\eta\partial (4+\eta\partial )}
on the hypercone reduces to $C_I=l(l-4)$.

To describe tensor conformal fields, we can consider homogeneous tensors
$O_{A_1...A_s}(\eta)$ on the hypercone. The irreducible representations of the conformal algebra are specified by the symmetric-traceless tensors which also
satisfy refs{\mack,\grillo},
\eqn\ten{\eqalign{\eta^{A_1}O_{A_1...A_s}(\eta)&=0\cr \partial^{A_1}O_{A_1...A_s}(\eta)&=0}}
Acting on these tensors, the Casimirs are not purely orbital, but they must be supplemented with an internal part, corresponding to a finite dimensional representation of O(4,2), $L_{AB}\delta_{A_1...A_s}^{B_1....B_s}+\Sigma_{A_1...A_s}^{B_1....B_s}$.

Unitarity imposes the following bounds \fer,
\eqn\bound{l\ge 1+j\qquad (j_Lj_R=0)}
\eqn\boundtwo{l\ge 2+j_L+j_R\qquad (j_Lj_R\ne 0)}
The two unitarity thresholds are satisfied by massless fields and conserved tensors field, respectively \grillo. The equations
\eqn\conserved{\partial^2\Phi_{(0,j)}=0}
\eqn\conservedtwo{\partial^{\alpha_1\dot\alpha_1}O_{\alpha_1..\alpha_{2j_L},\dot\alpha_1..\dot\alpha_{2j_R}}=0}
are indeed conformal covariant only if $l=1+j$ and $l=2+j_L+j_R$, respectively. This can be easily proved by considering the O(4,2) commutation rule $[K_\mu,P_\mu ]=-2i(g_{\mu\nu}D+M_{\mu\nu})$. 

In the AdS$_5$/CFT correspondence, all gauge invariant composite operators
in the  CFT can be associated with fields in AdS$_5$ \refs{\pol,\fer,\witten}. O(4,2)
is reinterpreted as the isometry group of AdS$_5$, and particles in AdS$_5$
are classified by the quantum number $(E_0,j_L,j_R)$ of the maximal compact
subgroup $O(2)\times O(4)$. In the identification with CFT operators, $E_0$
is identified with the scaling dimension and $(j_L,j_R)$ with the 4-dimensional Lorentz quantum number of the primary conformal operator. 

The O(4,2) covariant wave equation for a particle with quantum numbers $(E_0,j_L,j_R)$ in AdS$_5$ can be expressed in terms of the Casimirs of the conformal algebra. In this way, the mass of a particle can be expressed as a function of 
$(E_0,j_L,j_R)$. In the case of a scalar, for example, the Laplace operator in 
AdS$_5$ coincides with the quadratic Casimir $C_I$ and the wave equation reads,
\eqn\wave{\partial^2\Phi -\eta\partial (4+\eta\partial )=-{1\over 2}L_{AB}L^{AB}\Phi=E_0(E_0-4)\Phi,}
where we used eq. \tensor\ with $l\rightarrow E_0$. We see that the mass square of a scalar field in AdS$_5$ can be expressed in terms of the conformal dimension $E_0$ of the field
by $m^2=E_0(E_0-4)$.

In general we obtain the following relations \FFZ:
\item{--} Scalars: $\phi$
\eqn\scal{D(E_0,0,0):\qquad m^2=C_I=E_0(E_0-4)}
\item{--} Vector field: $A_{\mu}$
\eqn\vec{D\left (E_0,{1\over 2},{1\over 2}\right ):\qquad m^2=C_I=(E_0-1)(E_0-3)}
\item{--} Symmetric tensor: $g_{\mu\nu}=g_{\nu\mu}$
\eqn\tens{D(E_0,1,1):\qquad m^2=C_I-8=E_0(E_0-4)}
\item{--} Antisymmetric tensor: $A_{\mu\nu}=-A_{\nu\mu}$
\eqn\anti{D(E_0,1,0) \oplus D(E_0,0,1):\qquad m^2=C_I=(E_0-2)^2}

and for fermions,
\item{--} Fermions of spin $3/2$: $\psi_\mu$
\eqn\grav{D\left (E_0,1,{1\over 2}\right )+D\left (E_0,{1\over 2},1\right );\qquad m=E_0-2}
\item{--} Fermions of spin $1/2$: $\lambda$
\eqn\fermions{D\left (E_0,0,{1\over 2}\right )+D\left (E_0,{1\over 2},0\right );\qquad m=E_0-2}

Of particular relevance also in AdS$_5$ are the representations of O(4,2) which
saturate the unitarity bounds. The states which saturate the bound \bound\ in AdS$_5$ are called singletons and are topological fields living at the boundary
of AdS$_5$\refs{\fr,\frtwo,\frthree}. They cannot be associated with any gauge invariant operator in the CFT, but it is suggestive that they have the same quantum number of the conformal (generally colored) fundamental fields appearing in the CFT. In particular,
in the well known case of the duality between  type IIB on AdS$_5\times S^5$ and
N=4 SYM, the singletons in AdS$_5$ are in correspondence with the fundamental
N=4 SYM multiplet. All the other unitary representations in AdS$_5$ propagate
inside the bulk of AdS$_5$ and are in correspondence with gauge invariant composite operators in the CFT.

Let us discuss also the meaning of the second unitarity bound in AdS \boundtwo. This is the case of conserved currents in the CFT. The associated fields in
AdS$_5$ are massless fields sustaining a gauge invariance. In this way, the
global symmetries in CFT are associated with local symmetries in AdS$_5$ \fer. Here
are the simplest examples of this correspondence:
\item{--} stress-energy tensor/graviton $\qquad\qquad\qquad T_{\mu\nu}\rightarrow g_{\mu\nu}$
\item{--} global current/gauge field $\qquad\qquad\qquad\,\,\,\,\, J_\mu\rightarrow A_\mu$
\item{--} supercurrent/gravitino $\qquad\qquad\qquad\qquad\,\,\, 
 J_{\mu\alpha}\rightarrow \Psi_{\mu\alpha}$.

The conserved tensors in this list satisfy in CFT an equation like \conservedtwo. This garantees that the number of degrees of freedom contained in a conserved tensor matches with those of a massless particle in AdS$_5$. The number of
degrees of freedom of a conserved tensor is $(2j_L+1)(2j_R+1)-4j_Lj_R=2(j_L+j_R)+1$, which can exactly substain a representation of spin $(j_L+j_R)$ of the
little group O(3) of massless particles in the Poincar\'e limit of AdS$_5$.

In the case of representations for which $l> 2+j_L+j_R$, the fields in AdS$_5$ are massive, and the corresponding CFT operators are not conserved.
\newsec{The maximal supersymmetric case}
In the AdS/CFT correspondence, the conformal dimension $E_0$ is not in general
an integer. Only the conserved currents, in general, are protected under renormalization and have integer conformal dimensions; they are indeed associated with massless states in AdS. On the other hand, the generic CFT operator is expected to have
anomalous dimensions. Introducing supersymmetry in the game,
we have the notion of a ``chiral'' primary conformal operator, whose
dimension is not renormalized even if it is not a conserved tensor. In the 
familiar case of N=1 theories, the non-renormalization of the conformal dimension follows from the relation with the $U(1)_R$ charge ($E_0=q$). The notion of
chirality is associated with a shortening of the supersymmetric multiplet.

The Maldacena's conjecture \malda\ relates a SCFT in d spacetime dimensions with N extended supersymmetries with
the type IIB string (or M theory) on AdS$_{d+1}\times H$, where H is an Einstein manifold which gives rise, after dimensional reduction, to a 2N extended
gauged supergravity in AdS$_{d+1}$. 
Let us focus on the maximal supersymmetric case for d=3 or d=4. The conformal groups O(3,2) and O(4,2) are enlarged to the supergroups O(8/4) and SU(2,2/4),
with 32 conformal supercharges.

The superconformal algebra representations which saturate
the bound \bound\ correspond to the AdS$_{d+1}$ singletons, which are associated with the fundamental massless conformal fields defining the SCFT. The superconformal algebra representations which saturate the bound \boundtwo\ correspond to the {\it massless}
multiplets in AdS$_{d+1}$, associated with the CFT multiplet of global currents. Other fields besides the conserved tensors are, in general, required to close
a supersymmetric multiplet\foot{For example, three set of scalars in different representation of SU(4), fermions and tensor fields are required to close the N=4 SYM supercurrent multiplet \refs{\roo,\stelle} which contains the conserved stress-energy tensor, the
spinorial supersymmetry currents and the SU(4) R-symmetry currents.}; in the same way, the AdS$_{d+1}$ {\it massless} multiplet contains some massive fields,
in addition to the massless graviton, gravitinos and gauge fields \van. These massless multiplets are automatically short and their dimension is protected.
The massive multiplets can be short (chiral, with canonical dimensions) or long
(with anomalous dimensions). From the supergravity side, we know that all the KK states, coming from the dimensional reduction on H, are in short representations \nic\ and have integer dimensions. On the other hand, a generic string state has a mass which is not an integer and the corresponding CFT operator acquires anomalous dimension (in generally very large, in the limit in which supergravity can be trusted) \refs{\pol,\witten}.

 The superconformal representations can be induced by conformal primary fields.
A generic scalar superfield has $2^{16}$ components with spin range from 0 up to 4. The degeneracy of representations of the Lorentz group O(d-1,1) is Sp(16)
for d=3 and Sp(8) for d=4. Generic superfields correspond to representations of the Clifford algebra of O(32) where left and right representations are the bosons and fermions, respectively. 

Chiral primary superfields have $2^{8}\times l$ components, where l is the (finite) dimension of some representation of the Lorentz group and the R-symmetry
(O(8) for d=3, SU(4) for d=4).

Let us now focus on d=4. An unconstrained superfields, with $\theta^A_\alpha$ in the $(1/2,0)$ representation of SL(2,C) and N of SU(N), has     
generically $2^{4N}$ components, spanning the Clifford algebra of SO(4N)\stelle.
Bosons and fermions, corresponding to the even and odd powers in the $\theta$'s expansion, are the two chiral spinorial representations of O(4N). These superfields are extended to a representation of SU(2,2/N).

There are three types of different supermultiplets. 

The ultrashort representations correspond to singleton representations. These multiplets have degeneracy $2^4(2j_L+1)$ with spin range from $j_L-1$ to $j_L+1$\gunaydin. They correspond to a sequence of massless conformal fields with dimension  
and spin related as in eq. \bound. The N=4 SYM fundamental multiplet belongs to this class of representations, and it is special because it is self-conjugated.

Short representations correspond to massless or massive representations with degeneracy $2^8r$ where r is some finite dimensional representation of $SL(2,C)\times SU(4)$. Generic massless representations with spin range from $(j_L-1,j_R-1)$ to $(j_L+1,j_R+1)$ can be obtained by tensoring two singleton representations $(0,j_L)\times (j_R,0)$ giving $2^8(2j_L+1)(2j_R+1)$ states \gunaydin. The physical sector of these massless representations is obtained by having a gauge symmetry in AdS$_5$,  which reduces the number of components to $2^8(2(j_L+j_R)+1)$ \refs{\binegar,\fer,\FFZ}. These massless representations are obtained by a sequence of transverse conformal primary fields with dimension and spin satisfying eq. \boundtwo. 
Generically, a N-extended superfield corresponding to a short multiplet has an expansion in half of the $\theta$'s. It can be chiral or twisted chiral (for N=4). The spin range in the two cases is $(0,0)\rightarrow (N/2,0)$ or $(0,0)\rightarrow (N/4,N/4)$.
The superfields can be multiplied in a chiral way or in a non chiral way. In the first case one reproduces a superfield of the same structure, in the latter case one gets a long multiplet with $2^{4N}r$ components, with r the (finite) dimension of some representation of $SL(2,C)\times SU(N)$. The spin range of a long multiplet is $(\Delta j_L, \Delta j_R)=(N/2,N/2)$.

An example of long multiplet is the non-chiral multiplication of two short multiplet with $j_L=j_R=0$. This has as highest spin component a spin 4 singlet. This should be contrasted with  the massless graviton multiplet obtained by tensoring, in a (twisted) chiral way, two self-conjugate singleton multiplets (with $j_L=j_R=0$).

\newsec{Short multiplets}
 Let us consider in details the case of N=4 SYM. An abelian N=4 vector multiplet corresponds to the self-conjugate representation of SU(2,2/4).
The N=4 fundamental multiplet $W_{[AB]}(x,\theta,\bar\theta), A=1,..,4$ satisfies the constraints \stelle,
\eqn\constr{\eqalign{&W_{[AB]}={1\over 2}\epsilon_{ABCD}\bar W_{[CD]}\cr
&{\cal D}_{\alpha A}W_{[BC]}={\cal D}_{\alpha [A}W_{BC]}.}}
and contains, as first component, a set of six scalars $\phi_{[AB]}$ in the 6 of SU(4), which will be denoted also $\phi_l, l=1,..,6$. The superfield itself will be also denoted $W_l$.

$(x,\theta,\bar\theta)$ can be extended to an harmonic superspace \howe, where the superfield (now denoted W without indices) can be considered as a twisted chiral superfield \FA. In this way, all the product $W^p$ are still twisted chiral superfields and therefore are short multiplets\howe. In terms of the superfield defined in eq. \constr, we have $W^p=W_{\{l_1}...W_{l_p\}}$-traces \FA.

The superfield $TrW^2$ gives the supercurrent multiplet \refs{\roo,\stelle} and the massless graviton
in AdS$_5$ \gun. Notice that we are taking a trace in color space in order to get gauge invariant operators. The tower of superfields $TrW^p$ is the set of CFT
composite operators in short multiplets which is in one to one correspondence
with the KK states in AdS$_5\times S^5$ \FFZ. Using the explicit component expansion of W, given, for example, in \FFZ, and performing explicitly the superfield multiplication, we obtain the full spectrum of KK states computed in \refs{\van\guntwo}. The relation between masses and conformal dimensions of the CFT operators is that predicted by superconformal invariance and discussed in section 1  
(formulae \scal -\fermions ).

We can explicitly list  the operators in $W_p$ which are in a (0,p,0) SU(4) representation
and which therefore survive when fermions are neglected and only constant
values of the bosonic fields $\phi_l, F_{\mu \nu}$ are retained \FZL.

In terms of the singleton fields $\phi_l, F_{\alpha \beta}, F_{\dot \alpha \dot
\beta, }$ ($F_{\alpha \beta }=\sigma_{\alpha \beta}^{\mu \nu}F_{\mu \nu},
F_{\dot \alpha \dot \beta }=\bar F_{\alpha\beta}=\sigma_{\dot \alpha
\dot\beta}^{\mu \nu}F_{\mu\nu}$) we have,

\eqn\oneone{ \hbox{Tr}(\phi_{\{l_1}\cdots\phi_{l_{p}\}})-\hbox{traces}\qquad (0,p,0)}

\eqn\one{ \hbox{Tr}(\phi_{\{l_1}\cdots\phi_{l_{p-1}\}}F_{\alpha
\beta})-\hbox{traces}\qquad (0,p-1,0)}

\eqn\two{\hbox{Tr}(\phi_{\{l_1}\cdots\phi_{l_{p-2}\}}F_{\alpha
\beta}F^{\alpha \beta})-\hbox{traces}\qquad (0,p-2,0)}

\eqn\three{ \hbox{Tr}(\phi_{\{l_1}\cdots\phi_{l_{p-2}\}}F_{\alpha
\beta}F_{\dot\alpha \dot\beta})-\hbox{traces}\qquad (0,p-2,0)}

\eqn\four{\hbox{Tr}(\phi_{\{l_1}\cdots\phi_{l_{p-3}\}}F_{\alpha
\beta}F^{\alpha \beta}F_{\dot\alpha
\dot\beta})-\hbox{traces}\qquad (0,p-3,0)}

\eqn\five{\hbox{Tr}(\phi_{\{l_1}\cdots\phi_{l_{p-4}\}}F_{\alpha
\beta}F^{\alpha \beta}F_{\dot\alpha \dot\beta}F^{\dot\alpha
\dot\beta})-\hbox{traces}\qquad (0,p-4,0)}

We observe that SU(4) singlets are possible only up to p=4. There are exactly five
of them \FZL:

\eqn\singlets{\eqalign{
W^2\rightarrow &O_2={1\over 2}(F^2\pm F\tilde F)\qquad (\tilde F_{\mu
\nu}={i\over 2}\epsilon_{\mu \nu \rho \sigma}F^{\rho \sigma})\cr
W^2\rightarrow &T_{\mu \nu}=F_{\mu \rho}F_{\nu \rho}-
{1\over 4}\eta_{\mu \nu}(F_{\sigma\rho})^2\cr
W^3\rightarrow &O_3=F_{\mu
\sigma}F_{\rho \sigma}F_{\rho \nu}- {1\over 4}(F_{\sigma \rho})^2F_{\mu \nu}
\cr
W^4\rightarrow &O_4={1\over 4}[(F^2)^2-(F\tilde F)^2]=
F_{\mu \rho}F_{\nu \rho}F_{\mu \sigma}F_{\nu \sigma}- {1\over 4}(F^2)^2}}
In the AdS/CFT correspondence, these operators are seen, by analyzing the Born-Infeld D3 brane action, to couple to the s-wave of the type IIB (complex) dilaton, graviton, a self-dual combination of the NS-NS and R-R anti-symmetric tensors and a combination of the dilation mode of the internal ($S^5$) metric and the four-form anti-symmetric field with components on $S^5$ \refs{\KL,\pol,\das,\FFZ}.   

\newsec{Long multiplets}
The simplest example of long multiplet is easily constructed.
By tensoring two singleton we can obtain either a spin 2 multiplet (which is again a twisted chiral superfield) or a spin 4 multiplet. 
The six scalars have a product which decomposes as 20+1 under SU(4). The 20 are the first components of the spin 2 multiplet which corresponds to the massless
graviton in AdS$_5$. The singlet is the first component of a spin 4 multiplet which is not contained in the supergravity states in AdS$_5$, but should correspond to a massive string state \refs{\pol,\witten}.

The first component of this long multiplet is \stelle,
\eqn\long{Tr\phi_l\phi^l:\qquad Tr(W_{[AB]}W_{[CD]}\epsilon^{ABCD})|_{\theta =0}}
and can be roughly interpreted as (the non-abelian generalization of) the radial relative positions of the D3 branes in AdS$_5$, while the highest one is the spin 4, made with combinations of the following operators,
\eqn\spinfour{Tr(\phi^l{\cal D}_{\alpha_1}^\leftrightarrow\cdot\cdot\cdot{\cal D}_{\alpha_4}^\leftrightarrow\phi_l),\, Tr(F_{\mu\rho}{\cal D}_{\alpha_1}^\leftrightarrow {\cal D}_{\alpha_2}^\leftrightarrow F_{\nu\rho}),\, Tr(\bar\lambda^A\gamma_\mu{\cal D}_{\alpha_1}^\leftrightarrow{\cal D}_{\alpha_2}^\leftrightarrow{\cal D}_{\alpha_3}^\leftrightarrow\lambda_A)-({\rm traces})}
with dimension $E_0=6$. There are also spin (1,1) with $E_0=4$ in the 1+15+20+... of SU(4), and spin 1 with $E_0=3$ in the 1+15+... of SU(4).
In the free-field case, only the quoted representations appear.

This long multiplet is the N=4 embedding of the Konishi multiplet\konishi. In N=1
language, the Konishi multiplet is $\Sigma =S_ie^V\bar S_i$, where $S_i, i=1,..,3$ are the three chiral multiplets of the N=4 theory and V is the N=1 vector superfield. $\Sigma$ satisfies
\eqn\kon{\bar D\bar D \Sigma =W}
where W is the superpotential multiplet $W=g\epsilon_{ijk}f_{abc}S^{ia}S^{jb}S^{kc}$ where $f_{abc}$ are the structure constants of the gauge group. In the N=4 notations, the superpotential W is in the 10 of SU(4) since it is related to the gravitino mass term:
\eqn\massterm{W|_{\theta=0}\rightarrow {\cal W}_{[AB]}=gTr\phi_{[AB]}\phi^{[BC]}\phi_{[CD]}}
The $\theta^2$ terms in the N=4 superfield are,
\eqn\teta{\theta_\alpha^A\theta_beta^B\epsilon^{\alpha\beta}{\cal W}_{[AB]} +h.c. + \theta_\alpha^A\sigma^{\alpha\beta\mu\nu}\theta_\beta^BL_{\mu\nu} +
\theta_\alpha^A\sigma_\mu^{\alpha\dot\alpha}\bar\theta^{\dot\alpha}_B(J_{\mu A}^B + \delta^B_A t_\mu) + \cdot\cdot\cdot}
For an unconstrained superfield these 120 terms split into $(1/2,1/2)(15+1)+((1,0)+(0,1))(6)+(0,0)(10+\bar 10)$ i.e. 120=64+36+20. This is the $\theta^2$ term of a scalar superfield with $2^{16}$ components. In the free field case,
${\cal W}_{[AB]}\rightarrow 0$, $\partial^\mu J_{\mu A}^B=\partial^\mu t_\mu\rightarrow 0$ and we obtain a massless multiplet with $2^8\times 5$ components. Note that in free field theory there are additional conserved currents. This can be understood because in free N=4 Maxwell theory there is an additional SU(4) invariant fermionic current and two SU(4) currents which rotate independently scalars and fermions.
In fact, in the free-field limit, this corresponds to the statement that there are infinitely many massless representations in the product of two singleton representations \refs{\frtwo,\fer}. Since any irreducible representation which is contained in the product of two singletons is massless in AdS$_5$, in free field theory, the spin 4,7/2,3,5/2,2,3/2,1 would be conserved conformal fields. 
If the N=4 abelian multiplet is extended to a non-abelian SU(N) YM interacting multiplet, as given by eq. \kon,  then $Tr\phi_l\phi_m-{1\over 6}Tr\phi_p\phi^p$ is the first component of the graviton multiplet, while $Tr\phi_l\phi^l$ is the first component of a long massive spin 4 multiplet in AdS$_5$, which now contains all the $2^{16}$ components. Only in the abelian (free) case, this spin 4 multiplet become massless with only $2^8\times 5$ physical components.

In the free-field theory limit the Konishi multiplet corresponds to the $j_L=j_R=1$ massless spin 4 multiplet \gunaydin. In the interacting theory, the OPE of two chiral primary operators is expected to contain also the higher spin multiplets with $j_L=j_R=s/2$ (s=2,...). These massive representations will contain massive states with maximum spin $j=j_L+j_R+2=s+2$ and contain combinations of operators of the form:
\eqn\spinfourtwo{Tr(\phi^l{\cal D}_{\alpha_1}^\leftrightarrow\cdot\cdot\cdot{\cal D}_{\alpha_{s+2}}^\leftrightarrow\phi_l),\, Tr(F_{\alpha_1\rho}{\cal D}_{\alpha_2}^\leftrightarrow\cdot\cdot\cdot {\cal D}_{\alpha_{s+1}}^\leftrightarrow F_{\alpha_{n+2}\rho}),\, Tr(\bar\lambda^A\gamma_{\alpha_1}{\cal D}_{\alpha_2}^\leftrightarrow\cdot\cdot\cdot{\cal D}_{\alpha_{s+2}}^\leftrightarrow\lambda_A)-({\rm traces})}
These conformal operators should correspond to string states in N=4 supermultiplets up to spin $j_{MAX}=s+2$.

In the free field theory limit, the multiplet is massless only if the dimension $l(j_L,j_R)$ of a given operator is $2+j_L+j_R$. For other conformal dimensions the multiplet is massive. Note that, for unconstrained multiplets the conformal dimension l ia an arbitrary real number satisfying $l> 2+j_L+j_R$. This is consistent with the fact that in N=4 SYM theory a generic long multiplet has anomalous dimension, eventually related, at strong coupling, to the stringy massive excitations \refs{\pol,\witten}.

\newsec{Conformal invariance constraints on the CFT Green functions}
The correspondence between N=4 SYM and type IIB string theory on AdS$_5\times S^5$ can be explicitly used to compute field theory Green functions
in the limit in which the $\alpha'$ and string loop corrections can be neglected and the string theory reduces to the classical supergravity \refs{\freed,\tse,\seib}. In the SYM theory this corresponds to the t'Hooft limit $N\rightarrow\infty$ and $x=g^2N$ fixed, when x is also large \malda. The supergravity therefore describes the strong coupling dynamics of the large N limit of the N=4 SYM. 
All the long multiplet, which are associated to string excitations, are predicted to have large anomalous dimension $h =x^{1/4}$ \refs{\pol,\witten}, and their contribution to OPE and Green functions disappears in the strong coupling limit. In this way, at strong coupling, the OPEs and the Green functions get contribution only from the short (KK) states, whose general form has been discussed in section 3. However, when the 1/x ($\alpha'$) corrections are included, we can expect contributions also from the long multiplets.

A long multiplet contributing to the YM supercurrent Green function is just the Konishi multiplet discussed in section 4. It is indeed known that, at weak coupling, the Konishi multiplet appears in the OPE of the supercurrent multiplet \anselmi. Higher spin multiplets, of the form discussed in section 4,  are expected to contribute to the OPE as well.
In N=1 notation, indicating with $J_{\alpha\dot\alpha}(z), z=(x,\theta,\bar\theta)$ the supercurrent, we have \anselmi
\eqn\coni{\eqalign{&J(z)J(z')={c\over (s\bar s)^3} +{\Sigma(z')\over (s\bar s)^{2-h/2}} + \cdot\cdot\cdot\cr
&\Sigma(z)\Sigma(z') ={c'\over(s\bar s)^{2+h}} +{\Sigma(z')\over (s\bar s)^{1+h/2}} + \cdot\cdot\cdot }}
where $s=x-x'+i\theta\gamma\bar\theta'$. Here $c={1\over 24}(3N_v+N_\chi )$ and
$c'=N_\chi$, where $N_v$ and $N_\chi$ are the number of vectors and chiral multiplets (including color multiplicities), respectively.

This free field result for c and $c^\prime$ does not receive corrections up to two loops. For a generic N=1 theory with $N_\chi$ chiral multiplets in the representation T of the gauge group with a superpotential $W= Y_{ijk}\phi^i\phi^j\phi^k$\foot{The indices i contain both the color and the flavor indices. Similarly, T is, in general, a reducible representation of the gauge group.}, the two loop value for c is \jack
\eqn\gec{c={1\over 24} (3N_v+N_\chi+N_v{\beta (g)\over g} -\gamma^i_i)}
where $\beta (g)$ is the one-loop beta function and $\gamma^i_j$ the one-loop
anomalous dimensions \jones,
\eqn\anomal{\gamma^j_i= {1\over 16\pi^2}\left ( {1\over 2} Y_{ikm}Y^{jkm} -2g^2C^j_i(T)\right )}
The two-loop value of $c^\prime$ was computed in \anselmi\ and reads
\eqn\siprime{c'=N_\chi +2\gamma^i_i.}
In the N=4 SYM ($Y_{ijk}\rightarrow gf_{abc}\epsilon_{ijk}$) we see that there are no corrections up to two loop to the free-field value of c and $c^\prime$. In general, c, which can be related to an R-current anomaly, can be proved to be not-renormalized at all orders \anselmi. However, the anomalous dimension of $\Sigma$ is not zero also for conformal invariant theories and reads, at the first perturbative order,
\eqn\DIM{h={3\over 16\pi^2}{Y_{ijk}Y^{ijk}\over N_\chi}}
For N=4 we have $h ={3\over 16\pi^2}x$
\anselmi, but this value is  corrected, at strong coupling, to $x^{1/4}$ \refs{\pol,\witten}, and the contribution of the Konishi multiplet to the OPE, as well as of all the other higher spin long multiplets which acquire the same anomalous dimension $x^{1/4}$, becomes subleading.

We can use conformal invariance to get information on the structure of the OPE and Green functions \grillotwo. We will give formulae valid for arbitrary space-time dimension d. Let us consider only the constraints coming from the O(d,2) algebra. Supersymmetry will further imply selection rules on the operators which may appear in a given OPE expansion.

Consider, for simplicity, the OPE of two
primary scalars A and B. On the hypercone, we can write the following expansion\grillotwo:
\eqn\onthehy{A(\eta)B(\eta^\prime)=\sum_{n=0}^\infty (\eta\cdot\eta^\prime)^{-{1\over 2}(l_A+l_B-l_n +n)}D^{nA_1...A_n}(\eta,\eta^\prime)O_{A_1...A_n}(\eta^\prime)}
using the pseudo-differential operator,
\eqn\pseudo{\eqalign{&D^{nA_1...A_n}(\eta,\eta^\prime,\partial^\prime)=\eta^{A_1}\cdot\cdot\cdot\eta^{A_n}D^{-{1\over 2}(l_A-l_B+l_n +n)}(\eta,\eta^\prime,\partial^\prime)\cr
&D(\eta,\eta^\prime,\partial^\prime)=\eta\cdot\eta^\prime\partial^{\prime 2} -2(\eta\cdot\partial^\prime )(1+ \eta\cdot\partial^\prime )}}
which is well defined when $\eta^2=\eta^{\prime 2}=\eta\cdot\eta^\prime=0$.

Let us consider the case of a scalar operator O (n=0). Using the previous formula, the contribution of all the descendants of a given primary operator O, of dimension l, can be re-summed \grillotwo
\eqn\OPE{\eqalign{&A(x)B(0)=\left ({1\over x^2}\right )^{(l_A+l_B-l)/2}{\Gamma (l)\over \Gamma ((l+l_A-l_B)/2)\Gamma((l-l_A+l_B)/2)}C^O_{AB}\times\cr
&\int^1_0 duu^{(l_A-l_B+l)/2-1}(1-u)^{-(l_A-l_B+l)/2-1}\,_0F_1\left(
l+1-d/2;-{x^2\over 4}u(1-u)\partial^2_x\right )O(ux) + ...}}
Here $\,_0F_1(\nu;z)=\sum_{h=0}^\infty {1\over h!}{\Gamma(\nu )\over\Gamma(\nu +h)}z^h$ ia a generalized hypergeometric function.

Using the previous formula for the complete contribution of a given scalar operator O to the OPE, we can obtain the contribution of O and all its descendents to a four-point function of scalars. 

Conformal invariance implies that an n-point Green's function depends on an arbitrary function of n(n-3)/2 parameters if
\eqn\para{{n(n-3)\over 2}\le nd - {(d+2)(d+1)\over 2}}
and nd-(d+2)(d+1)/2 otherwise.
The functional form of two and three-point functions is therefore completely fixed, while the four-point function depends on an arbitrary function of two conformal
invariant parameters:
\eqn\fourpoint{\eqalign{<0|A(x)B(y)&C(z)D(t)|0> = [(x-y)^2]^{-l_B}[(x-z)^2]^{-{1\over 2}(l_A-l_B+l_C-l_D)}\times \cr &[(x-t)^2]^{-{1\over 2}(l_A-l_B-l_C+l_D)}[(z-t)^2]^{-{1\over 2}(l_C+l_D-l_A+l_B)}f(\rho,\eta)}}
where
\eqn\inva{\rho = {(x-t)^2(z-y)^2\over (x-y)^2(z-t)^2},\qquad \eta ={(x-z)^2(y-t)^2\over (x-y)^2(z-t)^2}.}
We can analyse the contribution of a conformal scalar and all its descendents to the four-point function in the s-channel, by simply using twice eq. \OPE. We
obtain the formula \grillotwo:
\eqn\final{\eqalign{&f(\rho,\eta)= \Gamma (l)\eta^{{1\over 2}(l_A-l_B+l_C-l_D)}\rho^{-{1\over 2}(l+l_C-l_D)}{\Gamma (-{1\over 2}(l_A-l_B+l_C-l_D))\over \Gamma ((l-l_A+l_B)/2)\Gamma((l+l_A-l_B)/2)}\cr  
&F_4\left ({1\over 2}(l+l_C-l_D),{1\over 2}(l+l_A-l_B);l+1-{d\over 2};1+{1\over 2}(l_A-l_B+l_C-l_D);{1\over\rho};{\eta\over\rho}\right ) +\cr
&\left( {\rho\over\eta}\right )^{{1\over 2}(l_A-l_B+l_C-l_D)}[(l_A-l_B)\rightarrow -(l_A-l_B), (l_C-l_D)\rightarrow -(l_C-l_D)].}}
where $F_4$ is a double hypergeometric function.

For identical states and d=4, we obtain (up to a multiplicative constant)
\eqn\last{f(\rho,\eta)=\rho^{-{l\over 2}}F_4\left ({l\over 2},{l\over 2};l-1,1;{1\over\rho},{\eta\over\rho}\right ).}

This result may be relevant in relating the four-point graviton amplitude in AdS$_5$ to the boundary correlator of four stress-energy tensors. The tree-level supergravity result should corresponds, in the boundary CFT, to the exchange of chiral primary operators with canonical conformal dimension. The $\alpha^\prime$ (or 1/x) expansion \banks\  should receive contributions from the unknown
OPE coefficients of the chiral multiplets, which cannot be specified by simply using conformal invariance, and also from the long multiplets. In the case of a four-point function, the form of the Green function is not completely specified by conformal invariance. We can determine the unknown function $f(\rho,\eta)$ in the case of a long multiplet, and confront it with the one for the exchange of a chiral multiplet, by expanding eq. \final\ for large $l=x^{1/4}$.
Note that the function $f(\rho,\eta)$ depends only on x. All the N dependence
of the Green's function is encoded in the OPE coefficients.

\centerline{\bf Acknowledgements}
S. F. is supported in part by the DOE under grant DE-FG03-91ER40662, Task C, the NSF grant PHY94-07194, and by the ECC Science
Program SCI*-CI92-0789 (INFN-Frascati). We would like to thank D. Anselmi, L. Girardello, M. Gunaydin, D. Z. Freedman, C. Fr\o nsdal, M. Porrati and L. Rastelli for helpful discussions. One of us (S.F.) would like to acknowledge the Institute of Theoretical Physics at Santa Barbara, where part of this work was done.

\listrefs
\end